\documentclass[%
preprint,
showpacs,
amsmath,amssymb,
aps,
pre,
]{revtex4-1}
\usepackage{graphicx}
\usepackage{dcolumn}
\usepackage{bm}
\usepackage{color}
\usepackage{multirow}

\begin{document}

\preprint{APS/123-QED}

\title{Dynamical traps in Wang-Landau sampling of continuous systems: Mechanism and solution}

\author{Yang Wei Koh}
\email{patrickkyw@gmail.com}
\author{Adelene Y.L. Sim}
\email{adelenes@bii.a-star.edu.sg}
\author{Hwee Kuan Lee}
\email{leehk@bii.a-star.edu.sg}
\affiliation{Bioinformatics Institute, 30 Biopolis Street, \#07-01, Matrix, Singapore 138671}

\date{\today}

\begin{abstract}
We study the mechanism behind dynamical trappings experienced during Wang-Landau sampling of continuous systems reported by several authors. Trapping is caused by the random walker coming close to a local energy extremum, although the mechanism is different from that of critical slowing down encountered in conventional molecular dynamics or Monte Carlo simulations. When trapped, the random walker misses entire or even several stages of Wang-Landau modification factor reduction, leading to inadequate sampling of configuration space and a rough density-of-states even though the modification factor has been reduced to very small values. Trapping is dependent on specific systems, the choice of energy bins, and Monte Carlo step size, making it highly unpredictable. A general, simple, and effective solution is proposed where the configurations of multiple parallel Wang-Landau trajectories are inter-swapped to prevent trapping. We also explain why swapping frees the random walker from such traps. The efficacy of the proposed algorithm is demonstrated.
\end{abstract}
\maketitle


\section{Introduction}

Wang-Landau sampling (WLS) \cite{Wang01a,Wang01b} is increasingly becoming an important tool for simulating a wide range of slow-relaxing, glassy systems which are not tractable by more conventional techniques such as molecular dynamics (MD) or Monte Carlo (MC) sampling using Metropolis algorithm. To date, WLS has been applied to a wide variety of systems ranging from random bond and random field systems \cite{Wang01b,Fytas11,Theodorakis11}, atomic fluid \cite{Shell02} and clusters \cite{Poulain06}, liquid crystals \cite{Jayasri05}, polymers \cite{Poulain06, Poulain07,Seaton10} and proteins \cite{Nagasima07,Swetnam10,Jonsson11}, to lipid membranes in explicit solvent \cite{Vogel13,Gai13}. In MD and MC, one frequently encounters the problem that the system gets trapped in some local minimum due to the presence of large barriers on the energy landscape \cite{Faller03,Zhou05,Belardinelli07a,Swetnam10}. WLS (and related entropic sampling techniques \cite{Berg91,Berg92,Lee93}), on the other hand, circumvents this problem by sampling from the density-of-states (DOS) instead of the Boltzmann distribution. Since there are no energy barriers to overcome, the system readily moves back and forth between the high and low energy regions of phase space, thereby ensuring efficient sampling. Due to the effectiveness of such multicanonical techniques, related methods such as Replica-Exchange Multicanonical Algorithm \cite{Sugita00, Mitsutake03a, Mitsutake03b, Mitsutake13} and Statistical-Temperature Monte Carlo \cite{Kim06, Kim09} are very efficient for simulating systems with rough energy landscapes.

Although  WLS is initially believed to be free from the problem of slow relaxation, there is increasing evidence that WLS is subjected to another kind of slowing down when applied to continuous systems. In an early study, Brown and Schultess \cite{Brown05} reported that for the ferromagnetic Heisenberg model, WLS is expensive when sampling from rare configurations close to the ground state.  Jayasri et al. \cite{Jayasri05}, in their study of liquid crystals, reported that WLS becomes extremely slow even for moderately large systems because the system often gets stuck in certain regions of configuration space. In their study of polypeptides and Lenard-Jones clusters using WLS, Poulain et al. \cite{Poulain06} also reported similar trappings of their systems.

\section{Models and the Wang-Landau method}

To examine the nature of these trappings, we consider two systems. The first is the two-dimensional $L\times L$ square-lattice, frustrated $XY$ model,
\begin{equation}
H=-\sum_{\langle i,j\rangle}J_{ij}\cos(\theta_i-\theta_j),
\label{eq.01}
\end{equation}
where $\theta_i\in(-\pi,\pi)$ ($i=1,\dots,L^2$) are $XY$ spins, $\langle i,j\rangle$ denotes summation over nearest-neighbor pairs. Periodic boundary condition is used for both lattice dimensions. $J_{ij}$ are identical independent random variables sampled from a gaussian distribution with zero mean and unit variance. Single-spin MC moves are used. The second system is an 8-mer poly-alanine molecule in a medium with dielectric constant $\epsilon = 2$. We used the Amber99-bs0 force-field \cite{Perez07} and simulated using a modified version of the MOSAICS package \cite{MOSAICS}. Torsion angle moves were used. We choose these two systems as illustrations because they are simple enough but yet possess rough energy landscapes with many local minima. WLS studies of the poly-alanine molecule have also been reported by Poulain et al. \cite{Poulain06}. 

Simulations were performed using the standard WLS algorithm \cite{Wang01a,Wang01b}. In WLS, the quantity of interest is the density-of-states $g(E)$, or the number of all possible states at the energy $E$. Once $g(E)$ is known, the partition function can be calculated,
\begin{equation}
Z=\sum_{\mathrm{\{configurations\}}} e^{-E/k_{B}T}=\sum_E g(E)e^{-E/k_{B}T},
\end{equation}
where $k_{B}$ is the Boltzmann constant and $T$ is the temperature, and most thermodynamic quantities can be obtained from $Z$. WLS estimates $g(E)$ via a random walk in energy space. Trials for a state with energy $E_i$ to one with energy $E_f$ are accepted according to the transition probability,
\begin{equation}
p(E_i\rightarrow E_f)=\mathrm{min}\left(1,\frac{g(E_i)}{g(E_f)}\right).
\label{eq.transition_rule}
\end{equation}
If $g(E)$ is the true DOS, Eq. (\ref{eq.transition_rule}) obeys detailed balance. However, $g(E)$ is initially unknown, and one makes a guess for it. Whenever an energy $E$ is visited according to the transition rule Eq. (\ref{eq.transition_rule}), the DOS is updated as $g(E)\leftarrow f\cdot g(E)$ where $f$ is called the modification factor. The role of the modification factor is to make gradual refinement of the initially inaccurate DOS. One begins the simulation with a large $f$ (usually $f=f_0=e^1\approx2.71828$) and gradually lowers it, in stages, to 1. When $f=1$ is reached, detailed balance for Eq. (\ref{eq.transition_rule}) is recovered, and one obtains the true DOS for $g(E)$. In practice, one works with $\ln g(E)$ to prevent numerical overflow, and the update rule is $\ln g(E)\leftarrow \ln g(E) + \ln f$. In our simulations, we start from $\ln f_0=1$ and lowers the modification factor according to $\ln f_{i+1}\leftarrow \frac{1}{2}\ln f_i$ upon completing the $i$th stage of simulation.

In the original formulation of WLS, one computes an energy histogram and uses its `flatness' as a criterion to proceed to the next stage of simulation. However, as this criterion has been found to be quite arbitrary by several subsequent studies \cite{Zhou05, Yan03, Morozov07, Morozov09, Lee01}, we use the criterion discussed in \cite{Lee01} to lower $\ln f$. Another modification we adopted here is the one suggested by Zhou et al. \cite{Zhou06} that for continuous systems the DOS should be linearly interpolated if the energy falls between the centers of two energy bins. Hence in our analysis the DOS is piecewise linear. The case of piecewise constant DOS will be discussed at the end of this paper.

\section{Trappings during Wang-Landau Sampling}

\begin{figure}
\includegraphics[scale=0.5]{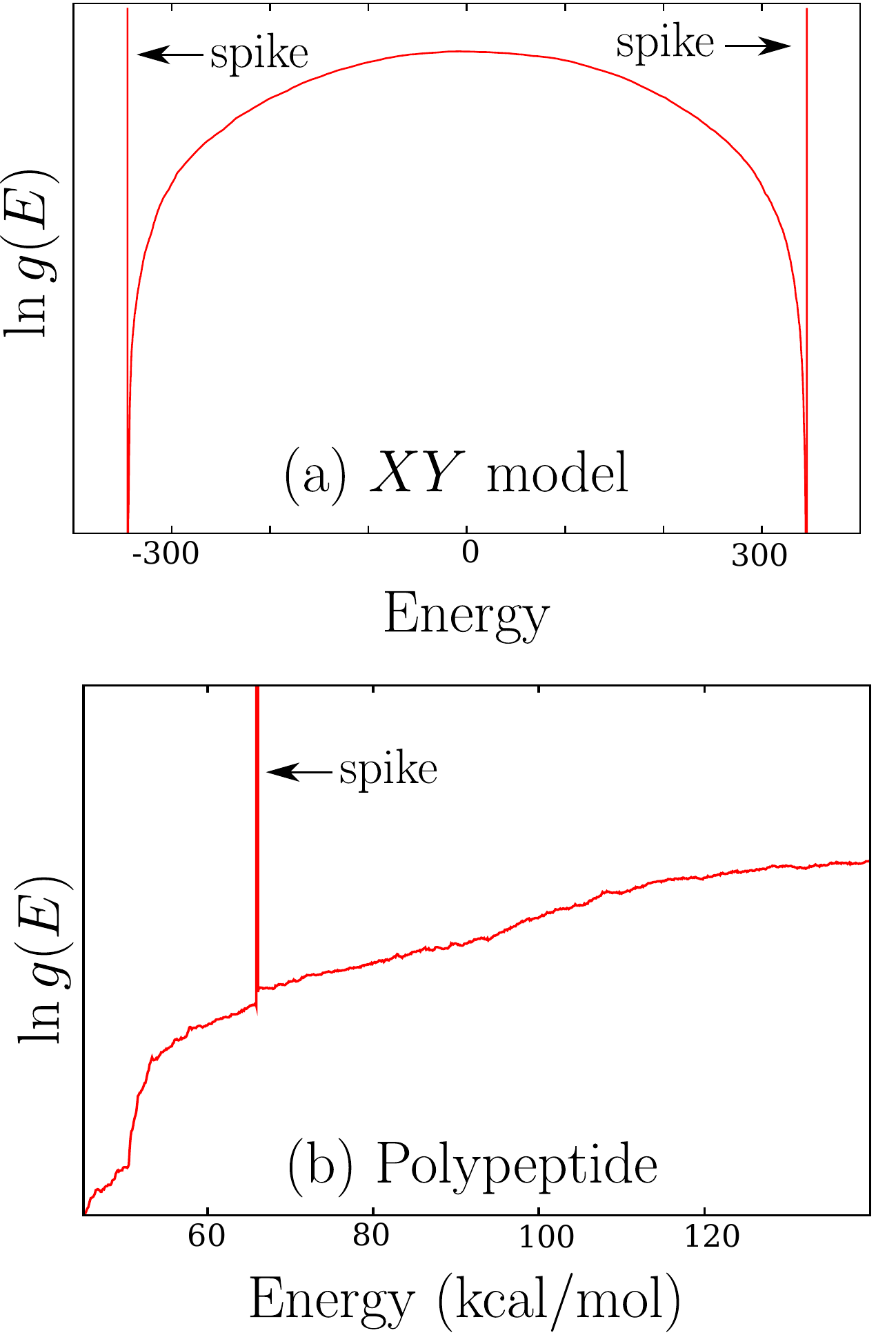}
\caption{(Color online) Spikes on the DOS for (a) the frustrated $XY$ model ($L=16$) and (b) the 8-mer poly-alanine. For both, we performed ten stages of WLS where each stage is simulated for $5\times 10^6$ updates per spin/angle, and the modification factor $\ln f$ is halved at at each stage. Shown are the DOS's from a single trajectory at the end of the tenth stage.}
\label{fig.spike}
\end{figure}

\begin{figure}
\includegraphics[scale=0.65]{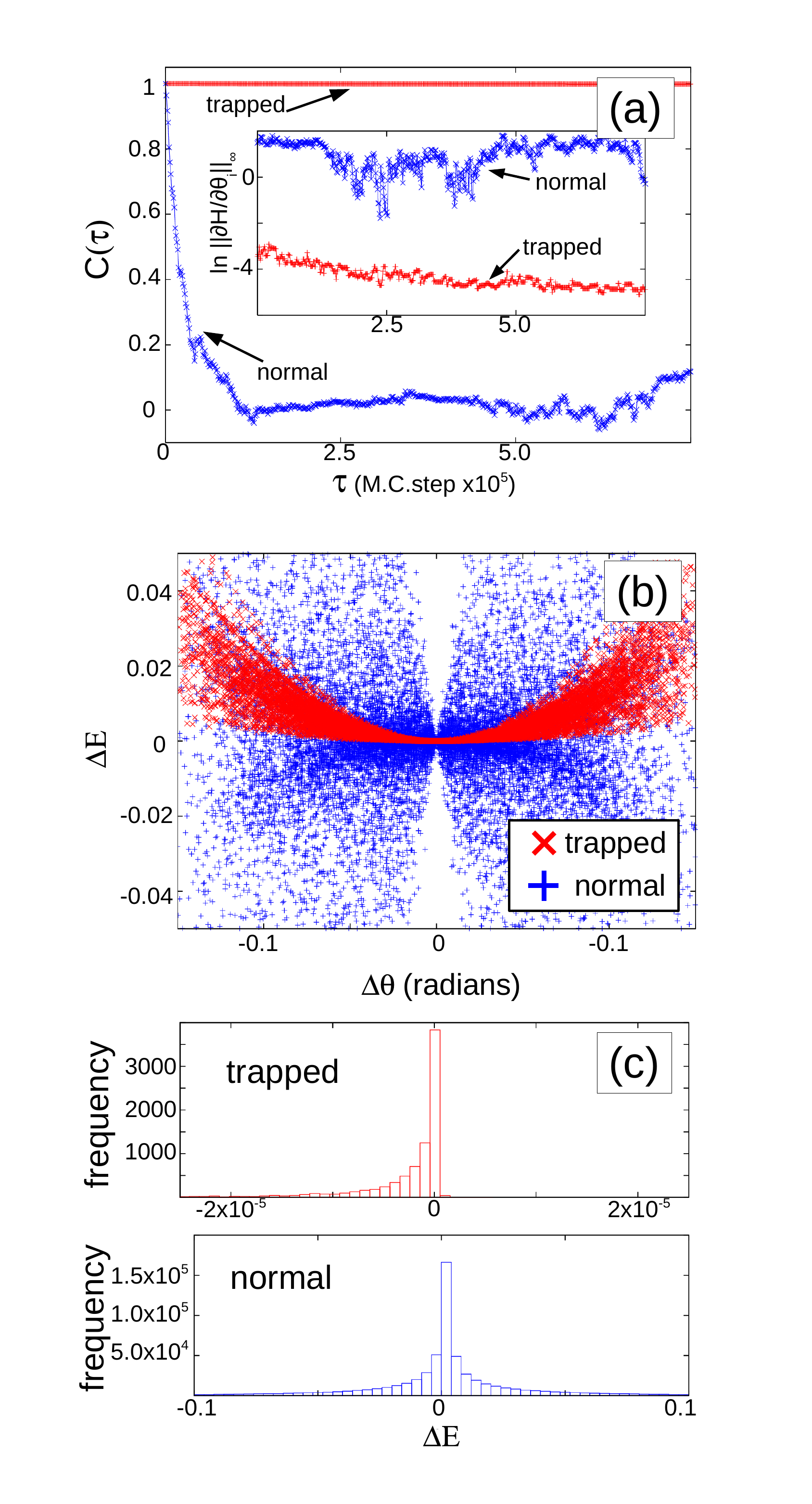}
\caption{(Color online) Comparing the behaviors between a trapped and normal segments of trajectory for the $XY$ model ($L=16$). (a) Autocorrelation function and infinity norm of force along the segments. (b) Scatter plot of proposed energy change $\Delta E$ versus proposed angle change $\Delta \theta$. (c) Energy distribution of accepted moves.}
\label{fig.trapped_vs_normal}
\end{figure}

Figs. \ref{fig.spike}(a) and (b) show examples of the DOS where the system exhibits signs of being trapped for the $XY$ model and poly-alanine molecule, respectively. The main feature is the high spikes, which are formed because the random walker stayed for an inordinate amount of time in a single energy bin, resulting in most of the modification factor $\ln f$ being accumulated there. We have checked that the system is capable of proposing new configurations with fairly large changes in energy with the MC step size we have used (c.f. Fig. \ref{fig.trapped_vs_normal}(b) and discussion below). Nevertheless, during the period of time when the random walker is stuck in a spiked bin, we found that all its MC moves result in small energy changes. The reason is not in a choice of small MC step size, but is due to the fact that the random walker has wandered close to a stationary point in configuration space. To ascertain that the random walker is close to a stationary point, we computed its energy gradient and found it to be very small indeed. 

Fig. \ref{fig.schematic} is a schematic diagram illustrating the mechanism of trapping and how a spike develops. Suppose, as shown in the inset, the system is at a local minimum with configuration $C$ and energy $E$. Let $E$ lie within the bin $b$ as shown. As the random walker is at a local minimum, the proposed energy $E^{\prime}$ must be $>E$, and the change in energy is small because the first derivative vanishes. Due to the steepness of the DOS, the acceptance ratio $\exp[\ln g(E)- \ln g(E^{\prime})]$ is very small, the move is mostly rejected, and $\ln f$ is added to the bin $b$. This process repeats itself, with the DOS at bin $b$ constantly increasing but the configuration hardly changing. If the system can propose a move to an energy, say $E^{\prime\prime}$ or higher, then the move may be accepted and the random walker can escape the trap. But this is highly improbable due to the vanishing of the gradient. The situation becomes more serious as system size increases because one usually increases the bin width in order to cover a wider energy range and reduce computational cost; however, the change in energy produced by single-spin or set of small torsion angle moves remains similar, and therefore becomes smaller relative to the increasing bin width. 

Figs. \ref{fig.trapped_vs_normal}(a)-(c) illustrate, for the $XY$ model ($L=16$), typical behavior of a trapped trajectory, compared against `normal' (i.e., not trapped) behavior. The trapped segment is taken from a continuous period of time the random walker is inside a spiked bin ($\approx8\times 10^5$ steps in this case). The normal segment is of similar length and taken from along the same trajectory but starting at a slightly earlier time before the random walker gets trapped. To monitor the configuration of the system, we computed the autocorrelation function,
\begin{equation}
C(\tau)=\frac{1}{L^2}\sum_{i=1}^{L^2}\cos\left[\theta_i(0)-\theta_i(\tau)\right],
\label{eq.02}
\end{equation}
where $\tau$ denotes the number of MC steps that have elapsed since the start of the segment. The results are shown in Fig. \ref{fig.trapped_vs_normal}(a). For the trapped segment, $C(\tau)$ remains close to unity throughout, meaning that its configuration hardly changes; for the normal segment, $C(\tau)$ decays to zero very quickly. The inset in (a) plots the component of the force $\partial H/\partial \theta_i $  with the largest magnitude (i.e. infinity norm) versus $\tau$. The force on the trapped segment is almost zero throughout. Fig. \ref{fig.trapped_vs_normal}(b) is a scattered plot showing the proposed energy change $\Delta E$ versus the proposed angle change $\Delta \theta$. As mentioned above, with the MC step size we are using, the random walker is able to, under normal conditions, propose new configurations with fairly large changes in energy, as shown by the large dispersion of blue points (i.e. normal segment). However, when trapped, $\Delta E$ is very small compared to the normal segment, and is distributed asymmetrically with very few energy-lowering moves. The acceptance rate is $\approx$1\% and $\approx$60\% for the trapped and normal segments, respectively. The energy distribution of the accepted moves are shown in Fig. \ref{fig.trapped_vs_normal}(c). For the trapped segment, the histogram is very narrow and asymmetric, with almost only energy-lowering moves being accepted. We also examined other cases of trappings, both for the $XY$ model (varying over a range of MC step sizes) and the alanine peptide molecule, and found the above behavior to be general in both systems. 

Spikes are usually formed during the early stages of the simulation when the DOS is still rough. A small random bump on the DOS like the one in Fig. \ref{fig.schematic} can serve as a `seed' for spike growth if the random walker accidentally visits a local minimum (or maximum, where similar arguments apply) whose energy happens to lie within the same energy bin as the bump.

Spikes do not necessarily have to form in the same bin everytime. This is because there are other local minima/maxima that reside in other energy bins. In addition, as shown in the inset of Fig. \ref{fig.schematic}, the random walker can sample other configurations, such as $D$, with the same energy $E$. This means that, spiking does not always occur in an energy bin where a spike was previously observed. Therefore spiking events are highly unpredictable.

A large amount of simulation time is wasted when the random walker is trapped because it spends less time sampling other parts of configuration space. When a spike is discovered at the end of one's simulation one should discard the results knowing it suffers from bad sampling, as shown later in our calculation of specific heat capacity. A simple fix to the problem would be to vary the energy bin width and/or sampling step size. Unfortunately, it is not possible to know \textit{a priori} the appropriate bin width and step size to choose to avoid spiking (if at all possible), unless inordinately large step sizes are used in conjunction with small bin widths. Even then, this is not a feasible general solution, as large moves make it difficult to sample low energy basins, while using small bin widths is impractical, especially for systems with large energy ranges.

\section{Proposed solution: Trajectory-swapping}

Here, we introduce a general solution. We propose running multiple trajectories of WLS in parallel, each starting from a different random seed (or initial condition). Periodically, one randomly pairs the configuration of each trajectory with a new DOS from another trajectory. If the random walker happens to be trapped in a spike, this swapping resets the configuration to a `safe' region of energy space and stops the spike from growing further. The idea is similar to that of replica-exchange Monte Carlo \cite{Hukushima96}, where a low temperature state trapped in a local minimum is swapped with a higher temperature one, enabling it to escape and sample other regions of configuration space. Swapping between different WL random walkers also ensures that the sampling is uniform in energy space. Our method is simple to implement and completely general, applicable to any energy binning or sampling step size.  The DOS from all the trajectories can be averaged at the end of the simulation so no computational resources are wasted.

Recently, Vogel et al. \cite{Vogel13} proposed a parallel WL method to extend WLS to large-scale problems. In their method, different random walkers in overlapping energy windows are run in parallel and occasionally exchanged so as to allow the entire energy range to be sampled simultaneously and quickly. The purpose of our current work, on the other hand, is to use parallelization and swapping as a way to prevent trapping. In addition, each of our random walker samples the entire energy range, and we do not define a separate transition probability for the exchange of two random walkers. The method of Vogel et al. may also be effective in spike prevention. However, the general focus and actual implementation of the two works are different.

We tested our algorithm on the frustrated $XY$ model. Let $\Theta^{\alpha}=(\theta^{\alpha}_1,\cdots,\theta^{\alpha}_{L^2})$ and $g^{\alpha}(E)$ denote, respectively, the configuration and DOS of the $\alpha$th trajectory, where $\alpha=1,\cdots,N$ denotes the random seed of each trajectory. Every $(T\times L^2)$th MC step, a shuffling algorithm randomly permutates the trajectory index $\alpha\rightarrow\alpha^{\prime}$, and assign $\Theta^{\alpha^{\prime}}$ as the new configuration of $g^{\alpha}(E)$.  The main factors affecting the efficacy of the algorithm are the time interval between swaps $T\times L^2$ and the total number of trajectories $N$. Swapping should be as frequent as possible so that spikes do not have the chance to build up. Using more trajectories reduces the probability of swapping two trajectories which are simultaneously trapped in the same energy bin. On the other hand, as we implemented the algorithm using message passing interface, the swapping process requires communication between different processors and incurs computational overhead, so one should also try to keep $T$ large and $N$ small.

As our swapping between trajectories is accepted with probability 1, strictly speaking, it violates the transition rule of WLS. However, as the number of swaps we performed is very small compared to the entire length of the trajectory, no serious error is introduced, as shown by our numerical experiments below. 

\section{Numerical simulations}

In our simulations, we consider one set of $J_{ij}$ for each $L$ \cite{self-averaging}, and study the effects of $T$, $N$, and random shuffling on the prevention of spike occurrences. The energy binning for the DOS is chosen such that the bins are narrow enough to support the energy distribution $g(E)e^{-\beta E}$ near zero temperature. Such narrow bin widths easily induce spiking in the original WLS algorithm. Admittedly, the original WLS might not induce spiking for other choices of bin widths, but we did not perform an exhaustive study on this issue, due to the unpredictability of spiking, as previously discussed. All $N$ trajectories are launched from the same initial configuration. Each $\ln f$ stage is simulated for $5\times 10^6 \times L^2$ MC steps, and $\ln f$ is halved after each stage. We then counted the percentage of the $N$ trajectories that have spiked DOS after the tenth $\ln f$ stage.

The results are summarized in Table \ref{table.count_spikes}. For each $L$, we studied $N=8, 16, 32$ and $T=250, 1000, 5000, \infty$ where $T=\infty$ means no swapping (i.e. original WLS). Each entry in the table shows the percentage of spiked DOS in the order $L=16/32$. For $T=\infty$, more than half of the $N$ trajectories exhibits spiking.  For each of the rest of the $N$-$T$ entries, the percentage is averaged over five runs where each run uses a different seed for the random shuffler. As expected, large $N$ (32) and small $T$ (250) is the most effective in preventing spike formation. Between $N$ and $T$, the latter is more important in preventing spikes. We found that is it important to keep $T<1000$ because otherwise swapping is ineffective. Increasing $N$ has a comparatively more gradual albeit still significant effect in preventing spikes.

We checked the accuracy of the DOS of our swapping algorithm by computing the specific heat capacity per spin, $c_v$, for $L=16$. From the five runs of $N=32$, $T=250$, we took one run and continued the swapping simulation until $\ln f\approx 5.96\times 10^{-8}$ \cite{schedule}. The $c_v$ of all 32 trajectories were calculated and averaged, giving $\langle c_v\rangle$. The value of $\ln g(E)$ in any spiked bin is replaced by the average of its two neighboring bins. The same is repeated for $T=\infty$, and results of replica-exchange Monte Carlo calculations are used as benchmark. The results are shown in Fig. \ref{fig.Cv}(a). $T=250$ and replica-exchange agree very well, even at low temperatures as shown in the inset. On the other hand, the $\langle c_v\rangle$ of $T=\infty$ suffers from large fluctuations. We found that the trajectories of $T=\infty$ can be separated into two types: (i) Those with no spikes in their DOS give a $\langle c_v\rangle$ that have successfully converged, as shown by the red (labelled `converged') curve in Fig. \ref{fig.Cv}(b). (ii) Trajectories with spikes on their DOS sometimes spend entire or even a few $\ln f$ stages confined either within the spiked bin itself or `bouncing' between the spike and edge of the DOS. When the random walker managed to escape, the $\ln f$ has already been lowered so much that the current modification factor does not give effective improvement to the DOS, and thus the DOS is effectively frozen at an earlier $\ln f$ stage. The $\langle c_v\rangle$ computed from these trajectories are therefore unconverged, as shown by the blue (labelled `unconverged') curve in Fig. \ref{fig.Cv}(b). Our observation also explains why the annealing method by Poulain et al. \cite{Poulain06} was successful. By re-increasing the $\ln f$, their random walker is allowed to redo those earlier $\ln f$ stages when it was trapped. 

\section{Discussions}



As mentioned above, the phenomenon of trapping is highly dependent on how the simulation is being set up, i.e. the kind of system, the energy binnings, the MC step size etc. We would like to highlight two scenarios where we think there is an increased likelihood of encountering trapping. The first is when it is preferable to use a small MC step size to achieve a good acceptance rate. In our study of the poly-alanine molecule, we found that the molecule sometimes fold into very tight conformations where small MC step sizes are necessary to enable the molecule to gently unfold (large moves will give rise to the `lever-arm' effect resulting in steric clashes). If small step sizes are used, then one needs to be aware of the trapping mechanism we have discussed above.

\begin{figure}
\includegraphics[scale=0.65]{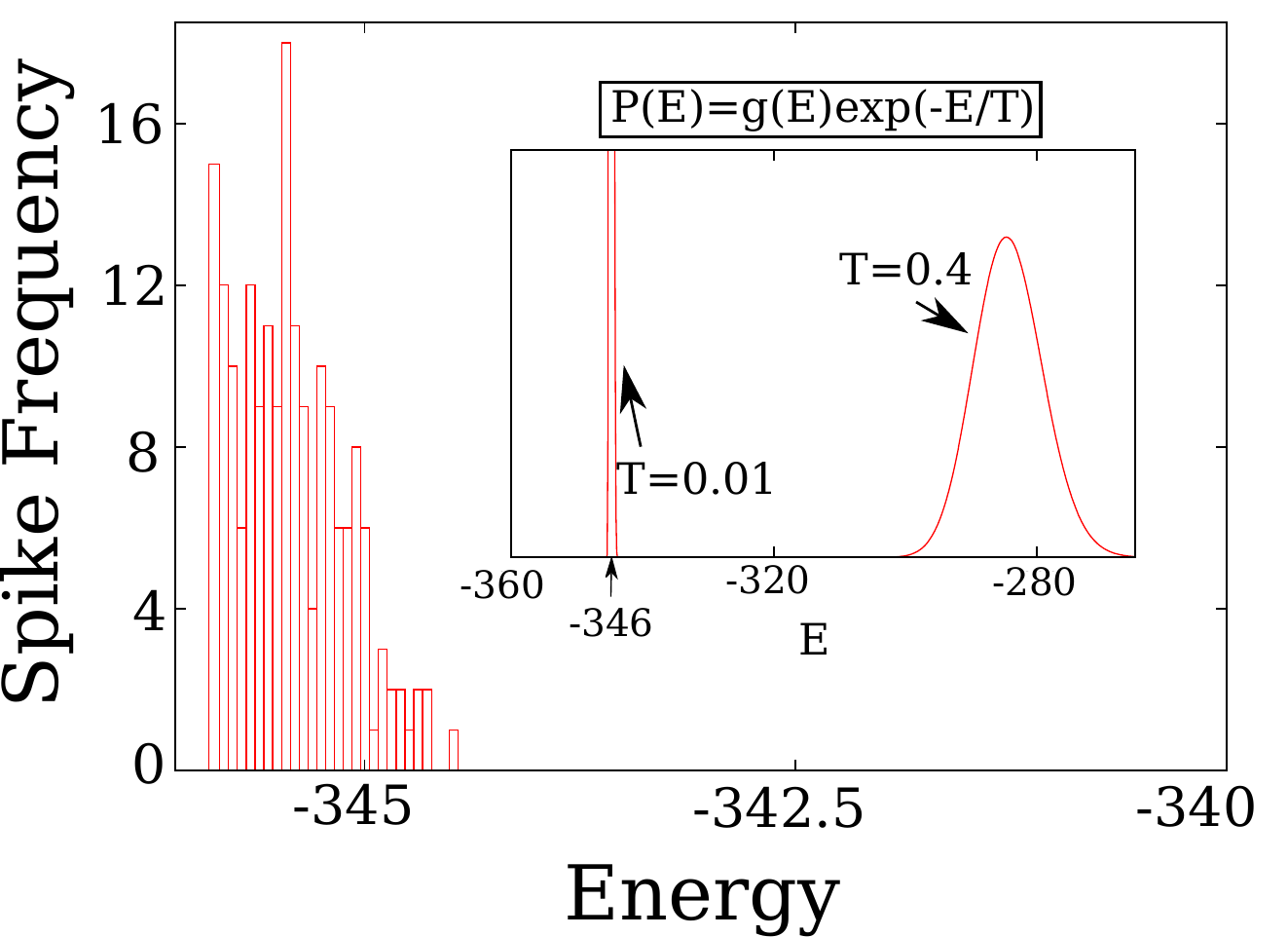}
\caption{(Color online) Distribution of spike locations with respect to energy for frustrated $XY$ model ($L=16$). Inset: The energy probability distribution $P(E)=g(E) e^{-E/T}$ at temperature $T=0.01$ (corresponding to $E\approx -346$ where spiking is dominant) and at $T=0.4$ (phase transition temperature).}
\label{fig.spike_distribution}
\end{figure}

The second scenario is when one is interested in low temperature behavior, which in WLS means sampling from low-lying energy states. To illustrate, we ran 200 independent, non-swapping trajectories for the $XY$ model ($L=16$) discussed above. At the end of the $15$th $\ln f$ stage, we checked if there is a spike in their DOS, and if yes, the energy bin at which the spike occurs. The energy distribution of the spike locations is shown in Fig. \ref{fig.spike_distribution}. Spiking occurs predominantly at an energy range near to the ground state at $E\approx-346$. This energy range corresponds roughly to the temperature $T=0.01$. In the inset of Fig. \ref{fig.spike_distribution}, the energy probability distribution $P(E)=g(E)e^{-E/T}$ at $T=0.01$ is shown, and we see that the distribution is supported on an energy region where spiking is dominant. Hence, one should be aware of the possiblity of trapping when one investigates low temperature phenomena.

\begin{figure}
\includegraphics[scale=0.65]{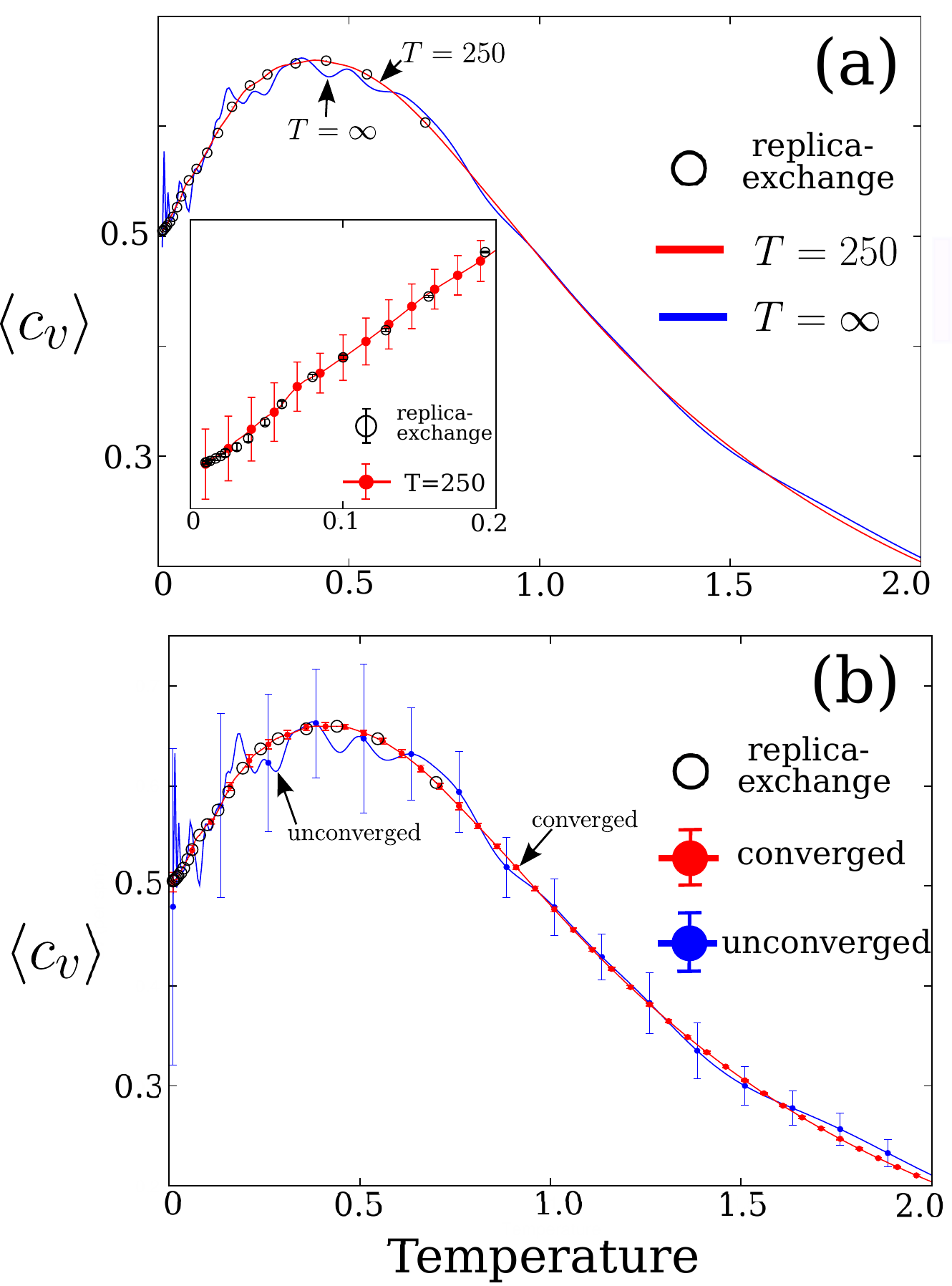}
\caption{(Color online) (a) Specific heat capacity per spin of the $XY$ model for a fixed set of $J_{ij}$ ($L=16$). The $\langle c_v\rangle$ for $T=250$ and $T=\infty$ is calculated by averaging over 32 trajectories. For replica-exchange Monte Carlo (circles), $\langle c_v\rangle$ is computed by averaging over 25 independent runs. Inset: Close-up view near zero temperature  of $T=250$ (filled circles) and replica-exchange (empty circles). Error bars indicate the standard deviation. The bars for replica-exchange are smaller than the circles. (b) For $T=\infty$. Converged: $\langle c_v\rangle$, averaged over 14 converged trajectories. Unconverged: $\langle c_v\rangle$, averaged over 18 unconverged trajectories. Error bars indicate standard deviation.}
\label{fig.Cv}
\end{figure}

On the other hand, trapping might not be so important when one studies high temperature behavior. We use the same $XY$ model as example. From Fig. \ref{fig.Cv}, we see that the transition temperature is around $T=0.4$. In the inset of Fig. \ref{fig.spike_distribution}, we plot the $P(E)$ at $T=0.4$. It is seen that the distribution lies quite far away from the energy region where spiking occurs. Hence, one might impose a low energy threshold to prevent the random walker from visiting low energy states if one is only interested in obtaining the transition temperature.

In our algorithm, swapping between trajectories occur with probability 1. As mentioned, this violates detailed balance. One way to rectify this is to treat the swapping as a MC step by itself. The acceptance probability for the swap has been proposed by Vogel et al. \cite{Vogel13},
\begin{equation}
P_{\mathrm{accept}}=\mathrm{min}\left[1, \frac{g_i(E[X])}{g_i(E[Y])} \frac{g_j(E[Y])}{g_j(E[X])}\right],
\label{eq:vogelswap}
\end{equation}
where $X$ and $Y$ are the configurations of trajectories $i$ and $j$ before the swap, and $E[X]$ and $E[Y]$ are their energies, respectively. $g_i(E[X])$ is the DOS of trajectory $i$ at energy $E[X]$. Hence, every few steps, a swap can be attempted and accepted according to Eq. (\ref{eq:vogelswap}). This way, detailed balance is obeyed and the swapping interval $T$ can be made quite small.

If the system under study is small enough for many multiple trajectories to be simulated on a single processor, one should indeed adopt Eq. (\ref{eq:vogelswap}). However, for large systems where one assigns different trajectories to different processors, one need to take into account the computational overhead coming from the communication among the processors during swapping. It may then be too expensive to use Eq. (\ref{eq:vogelswap}) to swap since frequent swapping will slow the computation down. On the rare occasion when a swap is attempted, it is best to let the swap succeed as we have done in this paper. Although theoretically less rigorous, our method is faster. Bouzida et al. has also shown within the context of the acceptance ratio method that occasional and infrequent violation of detailed balanced does not incur serious error \cite{Bouzida92}. Our calculation of the specific heat capacity also shows that there is no discernible error incurred from this violation of detailed balance during swapping.

\begin{figure}
\includegraphics[scale=0.5]{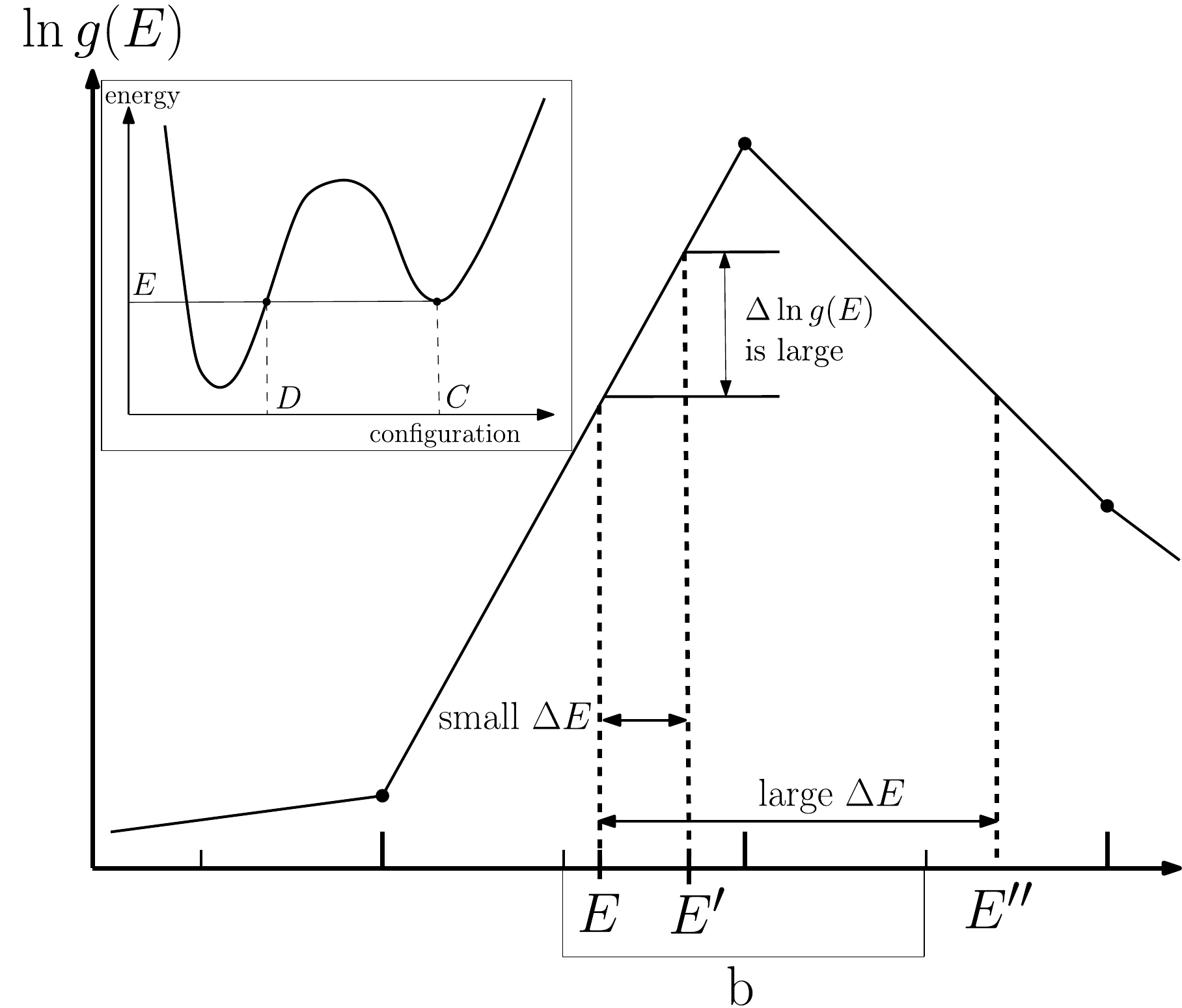}
\caption{Schematic diagram illustrating how the random walker gets trapped inside a spiked bin.}
\label{fig.schematic}
\end{figure}

Lastly, we discuss the case of piecewise constant DOS alluded to earlier. The original WLS algorithm takes the value of $\ln g(E)$ as constant within each energy bin. This will not give rise to any spiking, as without linear-interpolation the mechanism of Fig. \ref{fig.schematic} no longer holds. However, for piecewise constant DOS the energy bins at the very edge of the DOS will usually only be visited after the first few $\ln f$ stages have been completed. By the time the random walker drops into these bins, the difference in $\ln g(E)$ between these bins and the neighboring, higher-energy ones  is already so large that $\exp[\Delta\ln g(E)]$ is effectively zero, and the random walker stays a long time in these bins to let the DOS `catch up' before it can escape. This leads to another form of trapping. The swapping algorithm presented here is also applicable to this problem encountered by piecewise constant DOS.

We thank P. Poulain for useful discussions.

\begin{table}
\begin{tabular}{ccccccc}
\hline
\hline
       &  &    & \multicolumn{4}{c}{$T$}\\ \cline{4-7}
$N$  &  &    & 250 & 1000 & 5000 & $\infty$ \\ 
\hline
8 &   &   &  13(10)/0(0)  & 58(10)/5(6)  & 70(10)/48(20)  &  88/75 \\
16 &  &   &  9(8)/0(0)  & 49(10)/1(3)  & 64(10)/25(9)  &  69/69 \\
32 &  &   &  1(1)/0(0)  & 31(10)/0(0)  & 66(6)/38(10)  &  75/69 \\
\hline
\hline
\end{tabular}
\caption{Reducing spikes in the DOS by swapping. $N$: Number of trajectories. $T$: Number of MC steps (per spin) between swaps.  Each $N$-$T$ entry is ordered $L=16/32$, and shows the percentage of spiked DOS averaged over five different runs of the random shuffler. Parenthesis indicates the standard deviation. For $T$=$\infty$ (no swapping) only one run was performed for each $N$ and $L$.}
\label{table.count_spikes}
\end{table}

\begin{thebibliography}{99}
\bibitem{self-averaging} Although one should in principle average the results over the distribution of $J_{ij}$, as the system is approximately self-averaging at the large $L$ we are considering, we choose instead to focus calculations on other aspects.
\bibitem{schedule} The $\ln f$ schedule is as follows. From $\ln f=(0.5)^{0}$ to $(0.5)^{13}$, $5\times 10^6$ MCS per spin; $(0.5)^{14}$ to $(0.5)^{17}$, $4\times 10^7$; $(0.5)^{18}$ to $(0.5)^{20}$, $10^8$; $(0.5)^{21}$ to $(0.5)^{22}$, $2\times 10^8$; $(0.5)^{23}$ to $(0.5)^{24}$, $4\times 10^8$.
\end{thebibliography}
\end{document}